\documentclass[12pt]{article}
\usepackage{graphicx}
\begin{document}

\title{Stern-Gerlach Entanglement in Spinor Bose-Einstein
Condensates.}

\author{S. Cruz-Barrios$^{(1)}$, M. C. Nemes$^{(2,3\footnote{Permanent
Address.}\;\;)}$ and A. F. R. de Toledo Piza$^{(2)}$}

\maketitle

\begin{center}

{$^{(1)}$ Departamento de F\'{\i}sica At\'omica, Molecular y Nuclear,
Aptdo 1065, 41080 Sevilla \\ and Departamento de F\'{\i}sica Aplicada I,
Universidad de Sevilla \\ Sevilla, Spain}

{$^{(2)}$ Departamento de F\'{\i}sica--Matem\'atica, Instituto de
 F\'\i sica, Universidade de S\~ao Paulo, \\ C.P. 66318, 05315-970
 S\~ao Paulo, S.P., Brazil}

{$^{(3)}$ Departamento de F\'\i sica, ICEX, Universidade Federal de
 Minas Gerais,\\ C.P. 702, 30161-970 Belo Horizonte,
 M.G., Brazil}

\end{center}

\begin{abstract}

Entanglement of spin and position variables produced by spatially
inhomogeneous magnetic fields of Stern-Gerlach type acting on spinor
Bose-Einstein condensates may lead to interference effects at the
level of one-boson densities. A model is worked out for these effects
which is amenable to analytical calculation for gaussian shaped
condensates. The resulting interference effects are sensitive to the
spin polarization properties of the condensate.

\vspace{.2cm}


\end{abstract}

The classic Stern-Gerlach experiment\cite{sg} is currently understood
as providing for the spin analysis of a beam of particles using the
entanglement of spin and spatial degrees of freedom through the action
of inhomogeneous magnetic fields and subsequent spatial analysis of
the scattered flux\cite{HD}. In this paper we explore effects
resulting from this same type of entanglement when an essentially
static spinor Bose-Einstein condensate with large coherence length is
subjected to Stern-Gerlach field gradients. In the case of pure
condensates that are sufficiently dilute so that mean field effects
are a minor correction, the involved dynamics can be treated in a
rather straightforward and simple way for appropriate, non-trivial
field configurations using the ideas recently developed in
ref. \cite{sara}. 

A simple but still rather general model state for a pure spinor (spin
1 for definiteness) condensate is one in which all bosons occupy one
same single-boson spin-orbital $u(\vec{r},\sigma)$, where
$\sigma=-1,0,1$ is the spin variable. This state can be expressed in
terms of the creation operator

\[
a^\dagger\equiv\sum_\sigma\int d^3r\;u(\vec{r},\sigma)
\psi_\sigma^\dagger(\vec{r}),
\]

\noindent where $\psi_\sigma(\vec{r})$ is the spinor bosonic field
operator, as

\[
|\Psi^{(N)}\rangle=\frac{1}{\sqrt{N!}}\left(a^\dagger\right)^N
|0\rangle.
\]

\noindent This state is completely characterized by its one-boson
density, given by

\[
\rho(\vec{r},\sigma;\vec{r}\,'\sigma')\equiv\langle\Psi^{(N)}|
\psi_{\sigma'}^\dagger(\vec{r}\,')\psi_\sigma(\vec{r})
|\Psi^{(N)}\rangle=Nu(\vec{r},\sigma)u^*(\vec{r}\,',\sigma').
\]

\noindent The spin and position degrees of freedom are entangled
whenever the spin-orbital $u(\vec{r},\sigma)$ does not have the product
form $u(\vec{r})\chi(\sigma)$. A slightly more general form for a
non-entangled one-body density is

\begin{equation}
\label{obd}
\rho(\vec{r},\sigma;\vec{r}\,'\sigma')\rightarrow u(\vec{r}
)u^*(\vec{r}\,')\otimes\rho^{\rm (s)}(\sigma;\sigma')\equiv\rho^{\rm
(pos)}(\vec{r};\vec{r}\,')\otimes\rho^{\rm (s)}(\sigma;\sigma')
\end{equation}

\noindent where the second factor, representing the spin density
matrix, is allowed to be a mixed state. This information if of course
not sufficient to craracterize the underlying many-boson
state. Considering, as an example, a case in which $\rho^{\rm
(s)}(\sigma;\sigma')$ is diagonal, with eigenvalues $N_\sigma$ and
$\sum_{\sigma=-1}^1 N_\sigma=N$, both the pure $N$-boson state

\begin{equation}
\label{pst}
|\{N_\sigma\}\rangle=\frac{1}{\sqrt{N_1!N_0!N_{-1}!}}\prod_\sigma
\left(a_\sigma^\dagger\right)^{N_\sigma}|0\rangle,\hspace{1cm}
a_\sigma^\dagger\equiv\int d^3r\;u(\vec{r})\psi_\sigma^\dagger
(\vec{r})
\end{equation}

\noindent and the statistical mixture represented by the many-boson
density 

\begin{equation}
\label{mxst}
\sum_\sigma \left(a_\sigma^\dagger\right)^N|0\rangle\frac{N_\sigma}{N}
\langle 0|\left(a_\sigma\right)^N
\end{equation}

\noindent lead to that same one boson density.

The condensates recently produced using all-optical confinement
techniques \cite{chapman} may possibly be associated with one-boson
densities of type (\ref{obd}). This has been corroborated by applying a
Stern-Gerlach magnetic field to the condensate and observing its
splitting into three separate components. In general, the effect of
Stern-Gerlach fields is to generate spin-position entanglement, by
subjecting different spin components of the state to different boosts.
One may note, moreover, that the spin components are defined with
respect to a quantization axis associated to the field configuration
itself, and are therefore open to manipulation. In particular, a
mid-course change in the quantization axis will imply that the
position amplitude associated with each of the new spin components
will be a superposition of differently boosted components of the
initial state. This will give interference effects at the level of the
one-body density. Alternatively, a final spin selective observation
scheme implemented in terms of a quantization axis different from that
associated with the Stern-Gerlach field will also produce one-body
interference effects. Both of these schemes will be developed in what
follows. 

The mechanism responsible for the one-body interference can be
illustrated schematically in terms of a completely polarized
``travelling'' one-boson amplitude

\[
\phi(\vec{r},\sigma)=\frac{e^{i\vec{k}\cdot\vec{r}}}{\sqrt{V}}
\otimes|\sigma_z=0\rangle.
\]

\noindent If the field gradient is applied along the $x$-axis,
rotation of the state changes the spin factor to $(|\sigma_x=1
\rangle-|\sigma_x=-1\rangle)/\sqrt{2}$. The Stern-Gerlach boosts then
change the amplitude to (up to overall position independent phase
factors)

\[
\phi(\vec{r},\sigma)\rightarrow\frac{1}{\sqrt{2V}}\left(e^{i\vec{k}
\cdot\vec{r}+i\kappa x}|\sigma_x=1\rangle-e^{i\vec{k}\cdot\vec{r}-
i\kappa x}|\sigma_x=-1\rangle\right)
\]

\noindent where $\hbar\kappa$ is the momentum associated with the
boosts. The spin states $|\sigma_x=\pm 1\rangle$ can then be
re-expressed in the $z$ quantization axis as

\[
|\sigma_x=\pm 1\rangle=\frac{1}{2}\left(|\sigma_z=1\rangle\pm\sqrt{2}
|\sigma_z=0\rangle+|\sigma_z=-1\rangle\right)
\]

\noindent so that the spatial density associated with the
$z$-component $\sigma_z=1$ (say) is, after the Stern-Gerlach
entanglement

\[
\rho(\vec{r},\sigma_z=1;\vec{r},\sigma_z=1)=\frac{1}{2V}\sin^2\kappa x
\]

\noindent which shows interference fringes parallel to the
$z$-axis. Note, in particular, that this is independent of the initial
momentum $\vec{k}$ of the travelling amplitude. Eventually, an
analogy may be discerned between this mechanism and that involved in
Bragg diffraction experiments involving Bose-Einstein
condensates\cite{bragg1,bragg2}. More specifically, the coherent
superposition of amplitudes with different momenta resulting from the
action of the Stern-Gerlach field after subsequent redefinition of the
relevant quantization axis parallels that achieved through the action
of the Bragg pulses in the first ref.\cite{bragg1}.

We now turn to the description of the dinamics of the spinor
condensate under the field gradient. We assume that when the field is
applied the condensate is dilute enough so that effects of the
two-body interaction are negligible. The relevant dynamics is thus
contained in a second-quantized Hamiltonian of the form

\[
H=\sum_{\sigma\sigma'}\int d^3r\;\,\psi_{\sigma'}^\dagger(\vec{r})
\left[\frac{-\hbar^2\nabla^2}{2M}\delta_{\sigma\sigma'}+\langle
\sigma'|g\vec{S}\cdot\vec{B}(\vec{r})|\sigma\rangle\right]
\psi_\sigma(\vec{r}).
\]

\noindent The external magnetic field $\vec{B}(\vec{r})$ will be taken
to be such that there is a suitable Cartesian frame in which it has
the form, assumed to be valid throughout the spatial extention of the
condensate ($\sim 100\,\mu m$),

\[
\vec{B}(\vec{r})\rightarrow \vec{B}(x,z)=(B_0+b_1z)\hat{z}-b_1x\hat{x}.
\]

\noindent Typical values of the constants are $B_0\sim 50\;G$ and
$b_1\sim 2.5\;G\;cm^{-1}$.

As shown in ref \cite{sara}, the dynamics can be implemented in terms
of a set of semi-classical propagators associated to eigenvalues of
the projection of the spin operator onto the direction of the local
magnetic field. In view of the orders of magnitude given above, this
direction differs negligibly from the field $z$ axis. Therefore, if
this axis makes an angle $\theta$ with the quantization axis used to
express the spin density, in order to apply the semi-classsical
propagators one has to consider the rotated spin density

\[
\rho^{(\rm s)}(m;m')=\sum_{\sigma\sigma'}d^*_{\sigma m}(\theta)
\rho^{(\rm s)}(\sigma;\sigma')d_{\sigma' m'}(\theta)
\]

\noindent so that the complete initial density is given as

\[
\rho(\vec{r},m;\vec{r}\,'m')=\rho^{\rm (pos)}
(\vec{r};\vec{r}\,')\otimes\rho^{\rm (s)}(m;m').
\]

\noindent Propagation to time $t$ of each $m$ component is then
performed by

\[
\langle x_f,z_f|U_m(t)|x,z\rangle\approx\left(\frac{M}{2\pi i\hbar
t}\right)\exp\frac{i}{\hbar}{\cal{S}}_{\rm clas}
\]

\noindent where the classical action

\[
{\cal{S}}_{\rm clas}=\frac{M}{2t}[(x_f-x)^2+(z_f-z)^2]-\frac{1}{2}
gmb_1t(z+z_f)-\frac{(gmb_1)^2}{24M}t^3+gmB_0t
\]

\noindent has been evaluated using $m$-dependent trajectories in the
magnetic field. Note that $m$ is a constant of motion along the
trajectories due to the special choice of reference frame\cite{sara}.
The evaluation of the above expression involved neglecting terms of
the order $(b_1\Delta x/B_0)^2$, with $\Delta x$ of the order of the
propagation range. Since the evolution in the $y$ direction is free,
it has been ommited from the above expressions. The propagated density
at time $t$ is therefore given by

\[
\rho(\vec{r},m;\vec{r}\,'m',t)=\int d^3r_1\int d^3r_2
\langle\vec{r}|U_m(t)|\vec{r}_1\rangle\rho(\vec{r}_1,m;\vec{r}_2,m')
\langle\vec{r}_2|U_{m'}^\dagger(t)|\vec{r}\,'\rangle.
\]

\noindent The spin inclusive observed spatial density after
propagation in the Stern-Gerlach field is then given by the trace

\[
\rho^{\rm (obs)}(\vec{r},t)=\sum_m\rho(\vec{r},m;\vec{r},m,t).
\]

\begin{figure}
\centering
\begin{minipage}[c]{.45\textwidth}
\centering
\caption{Contour plots (sizes given in $\mu m$) of the spin inclusive
one-boson density (\ref{obd1}) integrated over $y$ for the shown
values of $t_1$ and $t$. The initial one-boson spin density is
diagonal with eigenvalue ratios $3:1:1$ for spin $z$-components
$-1,0,+1$ respectively. The initial gaussian amplitude has
$b_x=25\;\mu m$ and $b_{y,z}=15\;\mu m$. The angles $\theta$ and
$\theta_B$ are $\pi/2$ and $-\pi/2$ respectively. Magnetic field
parameters: $B_0=50\;G$ and $b_1=1.0\;G\,cm^{-1}$. Note that in (c)
interference fringes are too narow for accurate representation.}
\label{fig1}
\end{minipage}%
\hfill
\begin{minipage}[c]{.45\textwidth}
\begin{center}
\includegraphics[width=1.25in]{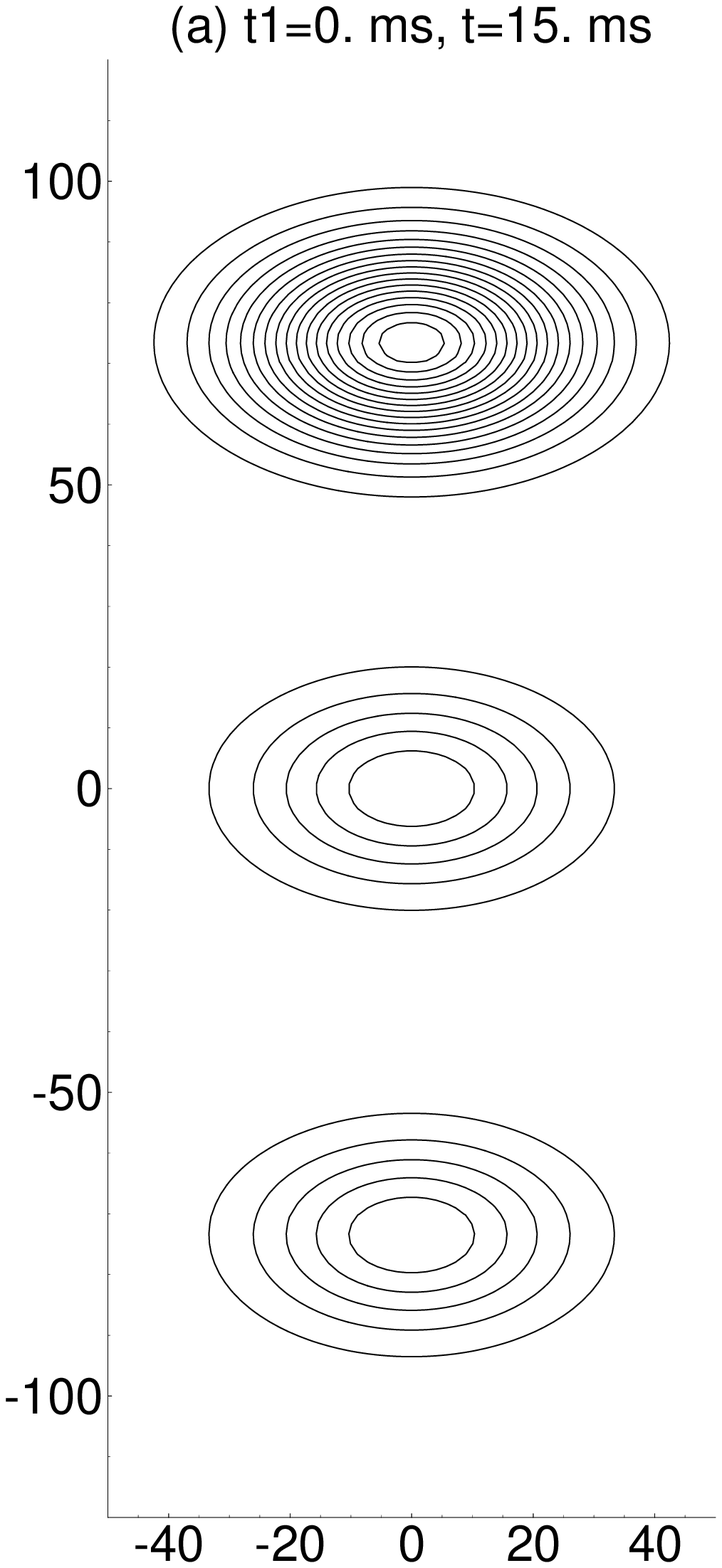}
\includegraphics[width=1.25in]{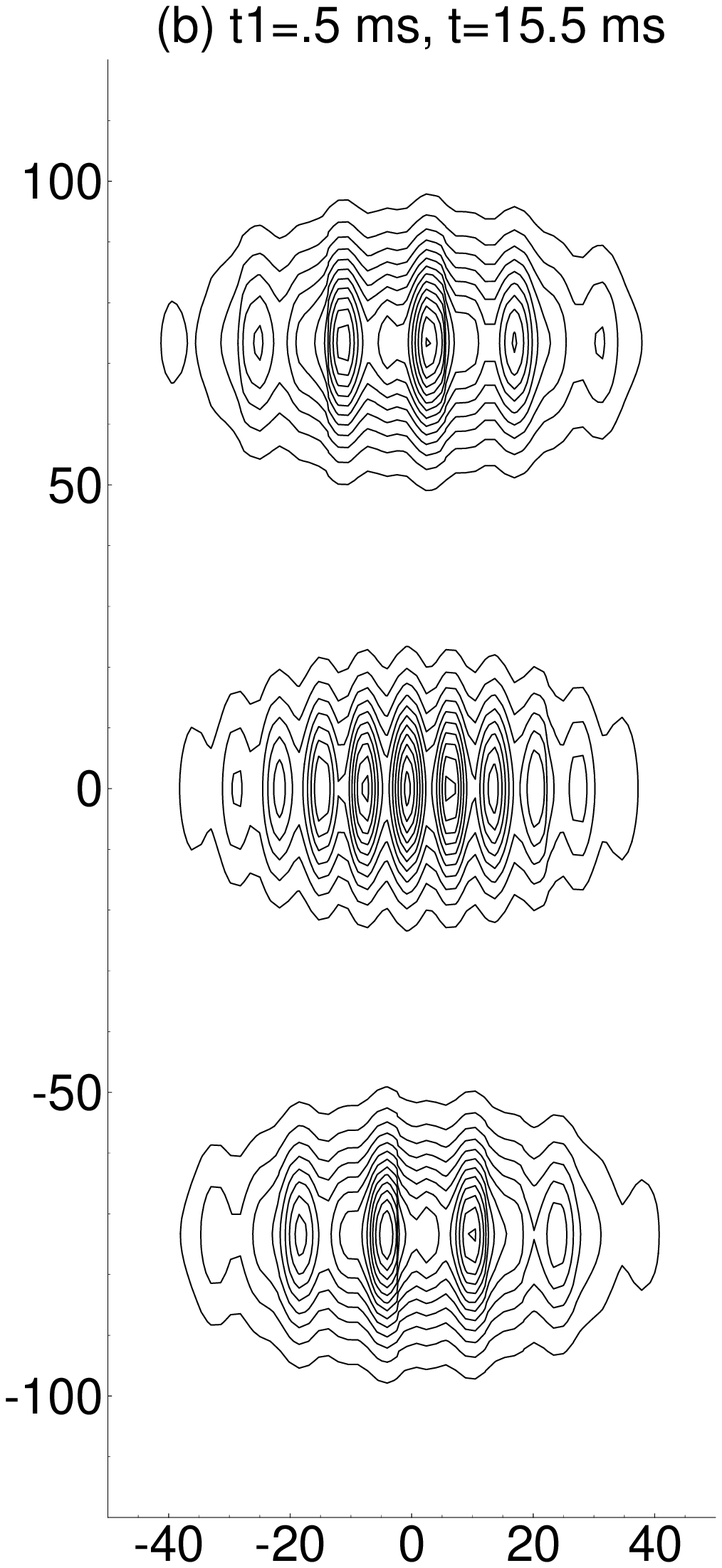}\\
\vspace{.2cm}
\includegraphics[width=2.5in]{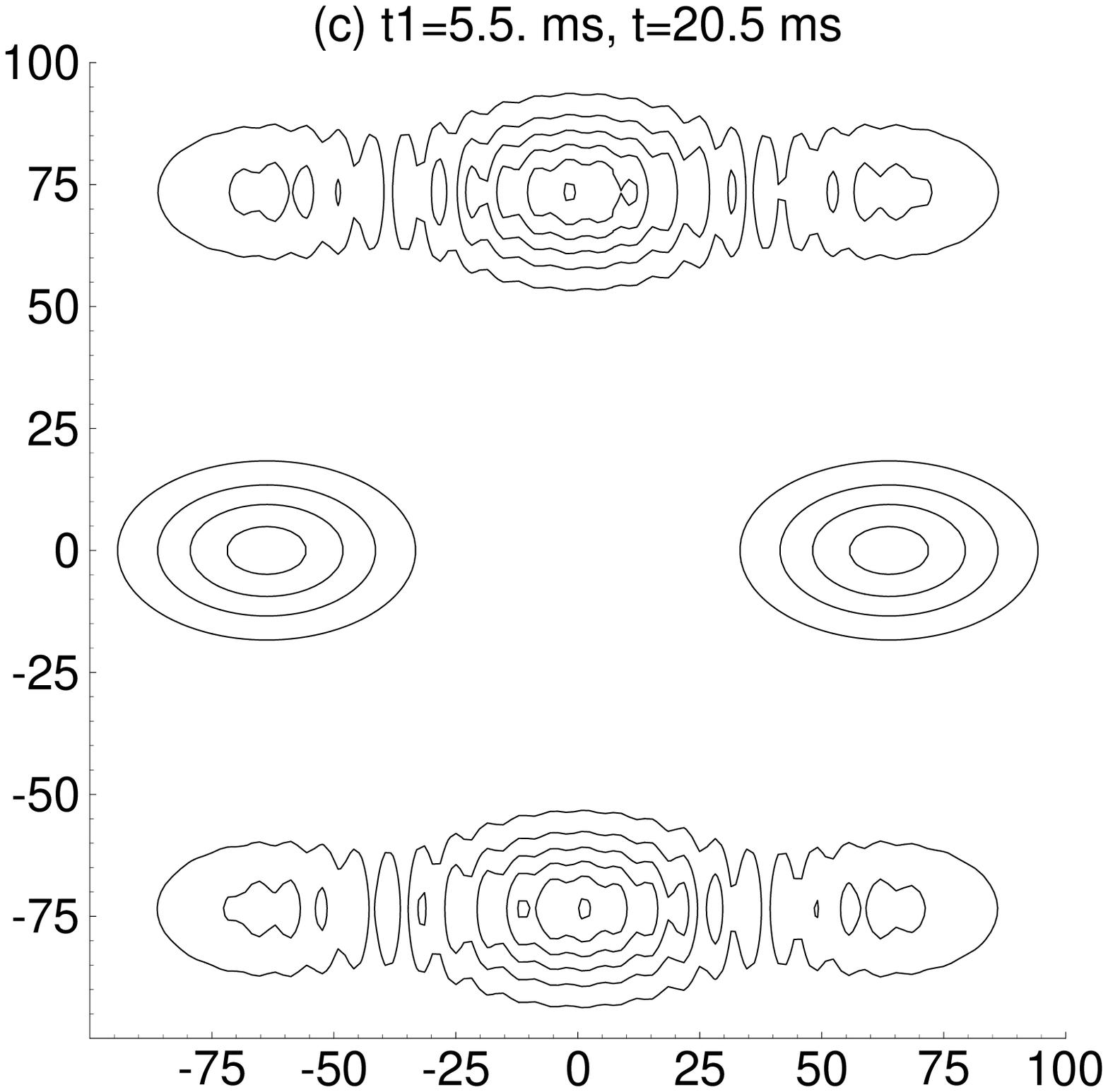}
\end{center}
\end{minipage}
\end{figure}

\noindent In the case of an initial spin density which is diagonal in
the magnetic field frame with weights $3/5:1/5:1/5$ for the components
$-1,0,+1$ and an initial gaussian spatial distribution one obtains the
result shown in fig. \ref{fig1}(a), for $t=15\;ms$ and field gradient
somewhat lower than the typical value quoted above. The resulting
distribution closely resembles that given in ref. \cite{chapman}.

We next consider a mid-course change of the Stern-Gerlach field. After
a first stage described by the above expressions and lasting for a
time $t_1$ a sudden rotation by an angle $\theta_B$ of $\vec{B}(x,z)$ is
performed around the $y$ axis. Strictly speaking, it is not important
that the field values are maintained, so that the rotation can be
achieved e.g. by superimposing an additional external field. The
initial condition for the second stage of the evolution will be then,
in the appropriate field frame

\[
\rho^{(2)}(\vec{r},m;\vec{r}\,'m',t_1)=\sum_{m_1m_2}d^*_{m_1m}(\theta_B)
\rho(\vec{r},m_1;\vec{r}\,'m_2,t_1)d_{m_2m'}(\theta_B).
\]

\noindent The propagation to $t>t_1$ is handled analogously to the
first stage, in terms of a propagator $\tilde{U}_m(t-t_1)$, leading to
a final density

\[
\rho^{(2)}(\vec{r},m;\vec{r}\,'m',t)=\int d^3r_1\int d^3r_2
\langle\vec{r}|\tilde{U}_m(t-t_1)|\vec{r}_1\rangle\rho^{(2)}
(\vec{r}_1,m;\vec{r}_2,m',t_1)
\langle\vec{r}_2|\tilde{U}_{m'}^\dagger(t-t_1)|\vec{r}\,'\rangle.
\]

\noindent The corresponding spin-inclusive final spatial density is
now

\begin{eqnarray}
\label{obd1}
\rho^{\rm (inc)}(\vec{r},t)&=&\sum_m\rho^{(2)}(\vec{r},m;\vec{r},m,t)=
\\ &=&\sum_m\sum_{m_1m_2}\int d^3r_1\int d^3r_2\langle\vec{r}|
{\cal{U}}_{mm_1}(t)|\vec{r}_1\rangle\rho(\vec{r}_1,m_1;\vec{r}_2,m_2)
\langle\vec{r}_2|{\cal{U}}_{m_2m}^\dagger(t)|\vec{r}\rangle\nonumber
\end{eqnarray}

\noindent where

\[
\langle\vec{r}|{\cal{U}}_{mm'}(t)|\vec{r}\,'\rangle=\int d^3r_1
\langle\vec{r}|\tilde{U}_m(t-t_1)|\vec{r}_1\rangle d^*_{m'm}
(\theta_B)\langle\vec{r}_1|U_{m'}(t_1)|\vec{r}\,'\rangle
\]

\noindent is the effective, spin non-diagonal propagator from time
zero to time $t$. Note that, during the first stage, different $m$
components are subjected to different boosts. After the rotation
$\theta_B$, the new $m$ components will in general be associated with
coherent superpositions of differently boosted spatial amplitudes,
leading to interferece patterns in the observed spin inclusive
one-boson density, as can be seen in fig. \ref{fig1}. This
interference will be absent in the special case in which the spin
density is diagonal in the reference frame defined by the field acting
in the first stage, due to the incoherent nature of the differently
boosted components.

\begin{figure}
\centering
\begin{minipage}[c]{.45\textwidth}
\centering
\caption{Contour plots (sizes given in $\mu m$) of the $m=+1$ spin
selected one-boson density (\ref{obd2}) integrated over $y$ for the
shown values of $t$.  The initial one-boson spin density is a pure
$s_z=-1$ state. The angles $\theta$ and $\theta_{\rm sel}$ are $\pi/4$
and $\pi/2$ respectively.  Gaussian and field parameters are the same
as in Fig. \ref{fig1}. Note that in (b) interference fringes are too
narow for accurate representation.}
\label{fig2}
\end{minipage}%
\hfill
\begin{minipage}[c]{.45\textwidth}
\begin{center}
\includegraphics[width=2.5in]{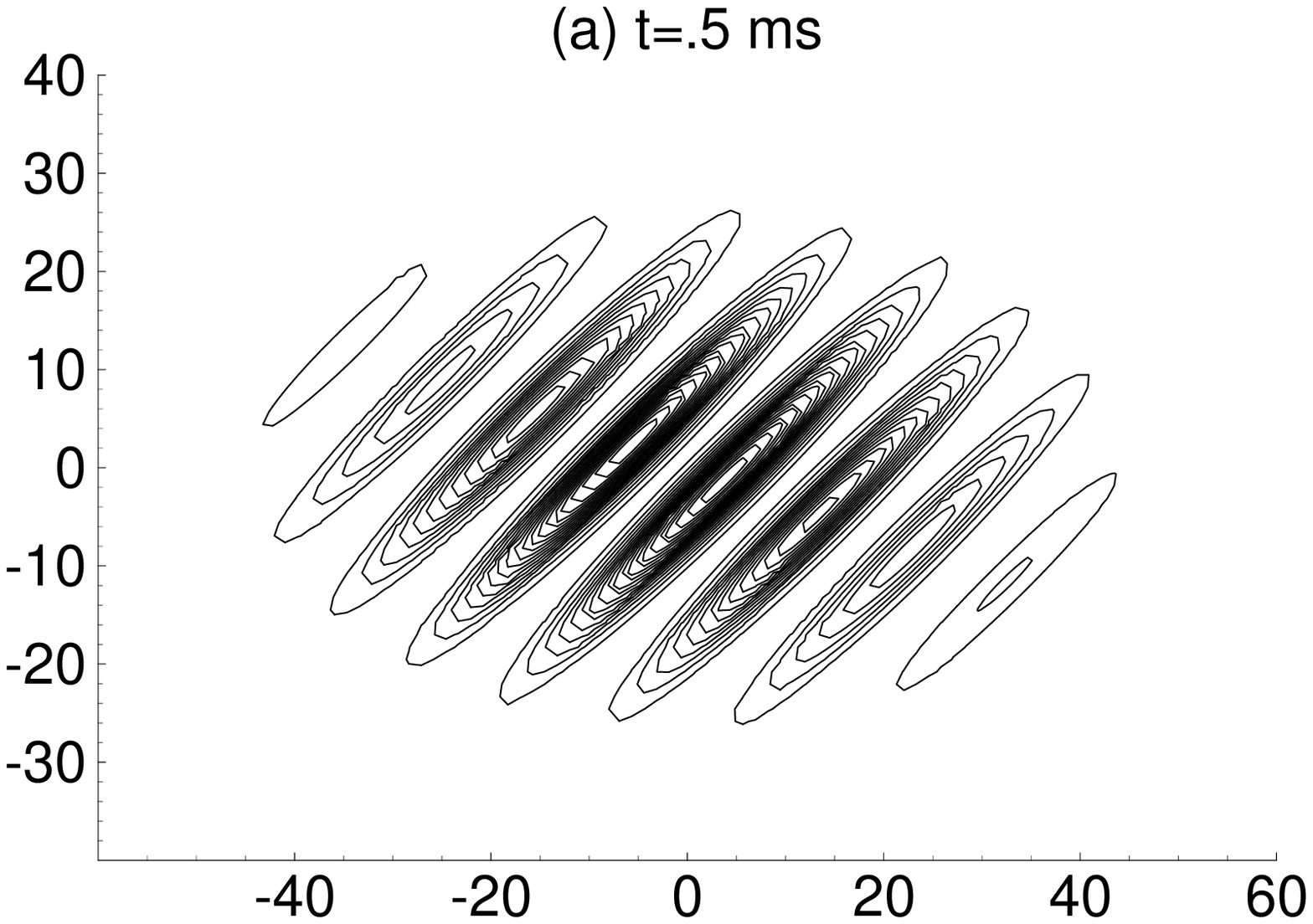}\\
\vspace{.2cm}
\includegraphics[width=2.5in]{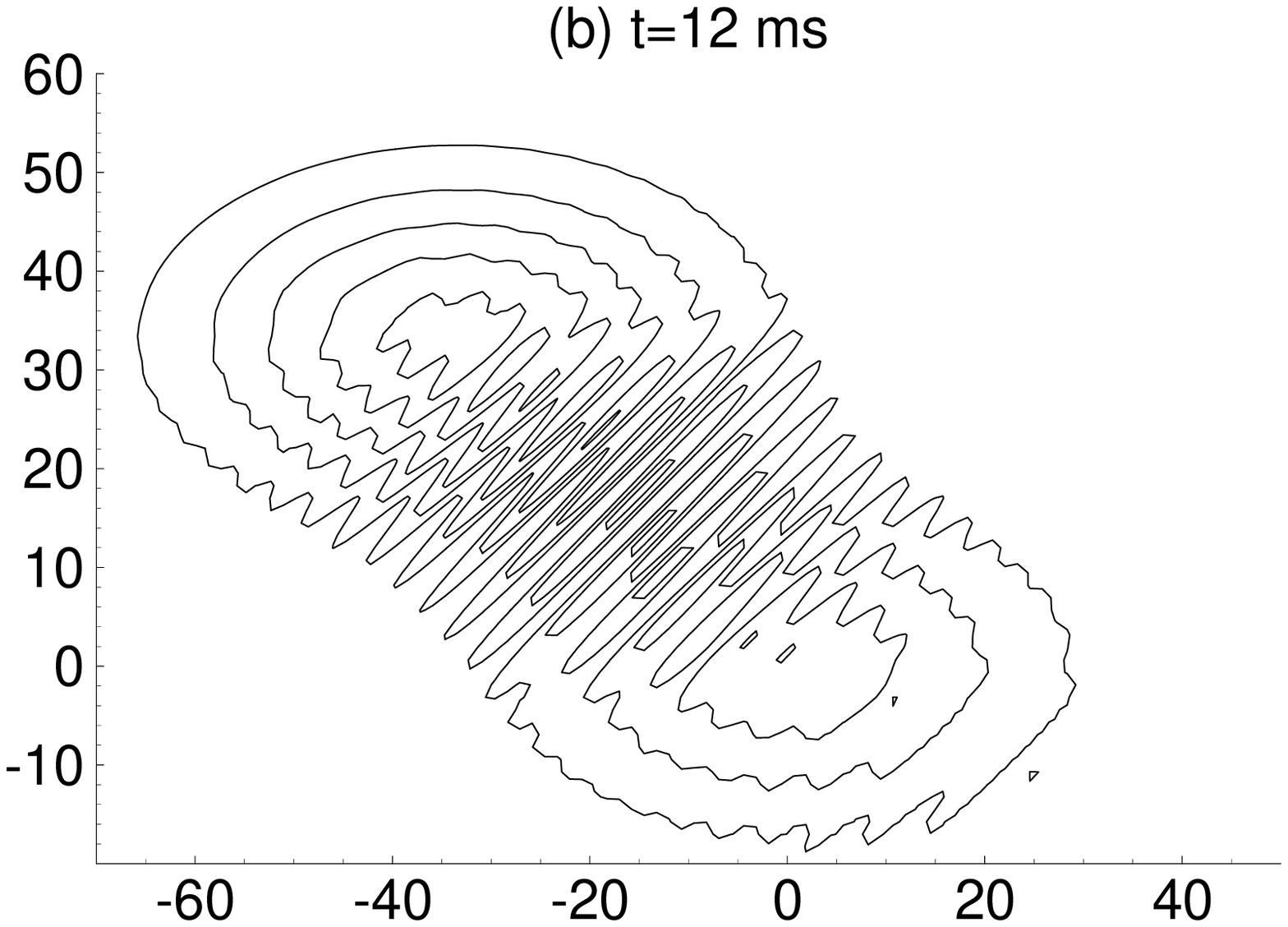}
\end{center}
\end{minipage}
\end{figure}

Alternatively, we may let the system evolve as in the first stage for
a time $t$, and then consider the resulting one-boson density
spin-selected along a quantization axis differing from the field $z$
axis by a rotation $\theta_{\rm sel}$ around the $y$ axis. In this
case the spatial one-boson density selected for the component $m$ of
the spin in the last frame will be given by

\begin{equation}
\label{obd2}
\rho^{\rm (sel)}(\vec{r},t)=\sum_{m_1m_2}d^*_{m_1m}(\theta_{\rm sel})
\rho(\vec{r},m_1;\vec{r},m_2,t)d_{m_2m}(\theta_{\rm sel}).
\end{equation}

\noindent which will also show interference effects for the same
general reasons discussed in the previous case. In particular, in the
special case in which the spin density is diagonal in the reference
frame defined by the field, no interference occurs in the one-boson
density. An example of this case is shown in fig. \ref{fig2}.

The initial (gaussian) spatial amplitudes leading to the patterns
shown in Figs. 1 and 2 have been taken as real, and therefore the
initial velocity field vanishes. The regime in which the two-body
interaction effects becomes negligible is in general reached after
release of the mean-field energy, which generates a non-vanishing
velocity field. The argument given above in order to illustrate the
one-boson interference mechanism indicates that this initial velocity
field does not change the interference pattern generated by the
Stern-Gerlach boosts. This has in fact been verified in the
calculations, by starting from narrower gaussians and letting them
spread freely for the appropriate time before applying the magnetic
field.

In both cases, the typical fringe spacing is determined by the de
Broglie wavelength associated with the relative velocities of the
interfering amplitudes. This depends both on the value of the magnetic
field gradient and on the duration of the first stage (in the first
case above) or of the exposure time to the field (in the second
case). For the $F=1$ states of $^{85}$Rb, fringes of the order of
several microns correspond to times of the order of $1\;ms$ or shorter
for $b_1\sim 1\;G\,cm^{-1}$.

It should be stressed that the above analysis has been carried out at
the level of the one-boson density. To the extent that laboratory
observation is sensitive to many-boson correlations\cite{jav},
further interference effects are to be expected. Notably, in the case
of the pure many-boson initial state (\ref{pst}) the Monte Carlo
procedure of ref. \cite{jav} will generate an interference pattern in
the extended $\vec{r},\sigma$ space even when the initial
Stern-Gerlach boosts are applied along the $z$-axis, which is however
invisible at the one-boson level. No many-body interferences occur,
under these conditions, in the case of the mixed many-boson state
(\ref{mxst}), due to the incoherece of the different spin
components. If, however, the initial Stern-Gerlach boost is applied
along a direction different from the $x$-axis, each one of the
incoherent, completely polarized spin components of the mixed
many-boson state will generate an interference pattern which seeps
down to the one-body level. The restriction to the one-body level
appears thus as a ``filter'' on the richer domain of many-body
effects. 

{\bf Acknowledgement.} S. C.-B. has been partially supported by the
Spanish CICyT, Project No. PB98-1111 and by the Pr\'o-Reitoria de
Pesquisa of the Universidade de S\~ao Paulo, and M.C.N. has been
partially supported by by the Funda\c{c}\~ao de Amparo \`a Pesquisa do
Estado de S\~ao Paulo (FAPESP) and by the Conselho Nacional de
Desenvolvimento Cient\'{\i}fico e Tecnol\'ogico (CNPq).

\newpage

\end{document}